\documentclass[preprint, review=false, screen=true,10pt]{acmart}

\usepackage{booktabs} 

\usepackage[ruled]{algorithm2e} 

\SetAlFnt{\small}
\SetAlCapFnt{\small}
\SetAlCapNameFnt{\small}
\SetAlCapHSkip{0pt}
\IncMargin{-\parindent}

\acmVolume{}
\acmNumber{}
\acmArticle{0}

\usepackage{etoolbox}
\makeatletter
\patchcmd{\maketitle}{\@copyrightspace}{}{}{}
\makeatother

\usepackage{natbib}
\setcitestyle{authoryear}
  
\begin{document}

\title{The 2D Tree Sliding Window Discrete Fourier Transform}  

\author{Lee F. Richardson}
\affiliation{%
  \institution{Department of Statistics and Data Science, Carnegie Mellon University}
  \streetaddress{5000 Forbes Ave}
  \city{Pittsburgh}
  \state{PA}
  \postcode{15213}
}
\author{William F. Eddy}
\affiliation{%
  \institution{Department of Statistics and Data Science, Carnegie Mellon University}
  \streetaddress{5000 Forbes Ave}
  \city{Pittsburgh}
  \state{PA}
  \postcode{15213}
}

\begin{abstract}
  We present a new algorithm for the 2D Sliding Window Discrete Fourier Transform (SWDFT). Our algorithm avoids repeating calculations in overlapping windows by storing them in a tree data-structure based on the ideas of the Cooley-Tukey Fast Fourier Transform (FFT). For an $N_0 \times N_1$ array and $n_0 \times n_1$ windows, our algorithm takes $O(N_0 N_1 n_0 n_1)$ operations. We provide a C implementation of our algorithm for the Radix-2 case, compare ours with existing algorithms, and show how our algorithm easily extends to higher dimensions. 
\end{abstract}


\thanks{Authors' addresses: Department of Statistics and Data Science, Carnegie Mellon University, 5000 Forbes Ave, Pittsburgh, PA 15213}

\maketitle

\renewcommand{\shortauthors}{Richardson, Eddy}

\section{Introduction}
In the early $19^{th}$ century, Joseph Fourier proposed that a function could be represented by an infinite series of sine and cosine waves with different frequencies, now called the Fourier transform. The Fourier transform burst into the digital age when \cite{cooley1965algorithm} re-discovered the Fast Fourier Transform (FFT) algorithm (see \cite{heideman1985gauss} for a history, which credits both the discrete Fourier transform (DFT) and FFT algorithm to Gauss). Since then, applications of the Fourier transform have soared (\cite{bracewell1986fourier}). 

The FFT reduces the number of operations for the DFT of a 1D length $n$ signal from $O(n^2)$ to $O(n \log{n})$. The FFTs computational savings are so significant that the Society of Industrial and Applied Mathematics (SIAM) listed the FFT as a top-ten algorithm of the $20^{th}$ century (\cite{cipra2000best}). That said, a downside of the FFT is that it operates on a global signal, meaning that local-in-time information is lost. Desiring locality, \cite{gabor1946theory} introduced a transform balancing time and frequency, which we call the Sliding Window Discrete Fourier Transform (SWDFT).

After \cite{gabor1946theory}, a plethora of researchers developed a variety of algorithms for the 1D Sliding Window Discrete Fourier Transform (1D SWDFT). Given the range of algorithms, we review the literature in Section \ref{sec:lit}, tying together previous work and major developmental themes. 

Today, Fourier transform applications extend beyond 1D. Like 1D, 2D Fourier transforms operate globally, but can capture local information using a 2D SWDFT. This paper presents a new 2D SWDFT algorithm, called  The 2D Tree SWDFT. Compared with existing algorithms, the 2D Tree SWDFT is fast, numerically stable, easy to extend, works for non-square windows, and is the only one with a publicly available software implementation. 

The rest of this paper is organized as follows. Section \ref{sec:lit} reviews existing SWDFT algorithms. Section \ref{sec:1dfswft} describes the 1D algorithm we extend in this paper. Section \ref{sec:2dfswft} derives our new algorithm, discusses implementation, software, and numerical stability, compares ours with existing algorithms, and shows how our new algorithm extends to higher dimensions. Section \ref{sec:discussion} concludes with a brief discussion. 

\section{Previous Work}
\label{sec:lit}
Since the $1960$s, researchers have produced two classes of algorithms for the SWDFT: recursive and non-recursive. Recursive algorithms update DFT coefficients from previous windows using both new data and the Fourier shift theorem, and non-recursive algorithms reuse FFT calculations in overlapping windows. This section briefly reviews the history of both algorithm classes, starting with recursive algorithms, then non-recursive algorithms, and concluding with recent developments in 2D. 

The first recursive SWDFT algorithm was introduced by \cite{halberstein1966recursive}, and has since been rediscovered many times (e.g. \cite{amin1987new,aravena1990recursive,lilly1991efficient,hostetter1980recursive,bitmead1982recursive,bongiovanni1976procedures}). Both \cite{sherlock1992moving} and \cite{unser1983recursion} gave 2D versions of the recursive algorithm, and \cite{sherlock1999windowed} derives recursive algorithms for different window functions. \cite{sorensen1988efficient} generalize the recursive algorithm to situations where the window moves more than one position, and \cite{park2014hopping} improved this generalization in an article titled ``The Hopping DFT''. \cite{macias1998efficient}, \cite{albrecht1997momentary}, and \cite{albrecht1999application} all proposed improvements to the recursive algorithm. Finally, the most cited recursive algorithm paper is \cite{jacobsen2003sliding}, likely because this paper provides an excellent description. 

One downside of the recursive algorithm is numerical error. In fact, \cite{covell1991new} proved that the variance of the numerical error is unbounded. Researchers responded by proposing numerically stable adaptations (e.g. \cite{douglas1997numerically,duda2010accurate,jacobsen2003sliding,park2015fast}). Although most of these adapted algorithms substantially increase computational complexity, \cite{park2017guaranteed} recently proposed the fastest, numerically stable recursive algorithm, called the ``Optimal Sliding DFT''. \cite{van2016constraining} recently reviewed recursive SWDFT algorithms. 

A numerically stable alternative to recursive algorithms are non-recursive algorithms. Whereas recursive algorithms update DFT coefficients from previous windows, non-recursive algorithms calculate an FFT in each window position, and reuse calculations already computed in previous window positions. Like the recursive algorithms, different non-recursive algorithms have been discovered by (at least) four different authors. \cite{bongiovanni1975non} proposed the first non-recursive algorithm, calling it the ``Triangular Fourier Transform'' (TFT). Shortly after, \cite{covell1991new} and \cite{farhang1992comment} proposed non-recursive algorithms, and \cite{farhang1995order} generalized the algorithm to arbitrary sized shifts. Since then, \cite{exposito1999fast,exposito2000fast} improved the algorithm for use in a digital relaying application. \cite{exposito1999fast} also pointed out that the butterfly diagram forms a binary tree, which was known (e.g. \cite{van1992computational}), but this was apparently the first time the tree was connected with the SWDFT. Recently, \cite{montoya2012efficient} and \cite{montoya2014short} gave further improvements to the non-recursive algorithm, including extension to the Radix-4 case. 

Independent of the non-recursive algorithms proposed by \cite{covell1991new,farhang1992comment,bongiovanni1975non}, \cite{wang2009meg} discovered another non-recursive algorithm while conducting a magnetoencephalography experiment. \cite{wang2009meg} capitalized on the binary tree structure of the Radix-2 FFT, and used it to derive a 3D data-structure shaped like a long triangular prism, with one binary tree for each window position. \cite{wang2010prfft,wang2012parallel} further developed a parallel version of this algorithm. This paper extends the algorithm described by \cite{wang2009meg}, which we call the Tree SWDFT, to 2D. 

Recently, researchers have proposed new algorithms for the 2D SWDFT. \cite{park20152d} extended the 1D recursive algorithm to 2D, and \cite{byun2016vector} proposed a 2D SWDFT based on the $2 \times 2$ Vector-Radix FFT algorithm (\cite{rivard1977direct,harris1977vector}). We use the algorithms of \cite{park20152d} and \cite{byun2016vector} as comparisons for the 2D Tree SWDFT algorithm proposed in this paper. 

\section{The 1D Tree Sliding Window Discrete Fourier Transform}
\label{sec:1dfswft}
This section describes the algorithm in \cite{wang2009meg}, which we extend in Section \ref{sec:2dfswft}. After defining the 1D SWDFT, we show a tree data-structure for the FFT, then illustrate how the tree data-structure leads to the 1D Tree SWDFT algorithm. 

  \subsection{The 1D Sliding Window Discrete Fourier Transform}  
  The 1D SWDFT takes a sequence of DFTs within each position of a sliding window. Specifically, let ${\bf x} = [x_0, x_1, \ldots, x_{N - 1}]$ be a length $N$ complex-valued signal, and let $n \leq N$ be the window size. Indexing the window position by $p = n - 1, n, \ldots N - 1$, the 1D SWDFT is: 

  \vspace{-1em}
  \begin{eqnarray}  
    \label{eq:1dswft}
    a_k^p &=& \frac{1}{n} \sum_{j = 0}^{n - 1} x_{p - n + 1 + j} \omega_n^{-jk}
  \end{eqnarray}
  \vspace{-1em}

  \hspace{-1.2em} for $k = 0, 1, \ldots N - 1$, where $\omega_n = \exp{\frac{i 2 \pi}{n}} = \cos{\frac{2 \pi}{N}} + i \sin{\frac{2 \pi}{N}}$. 

  A straightforward calculation of Equation \ref{eq:1dswft} takes $P n^2$ operations, where $P = N - n + 1$ is the number of window positions. Replacing the DFT with an FFT for each window position reduces this to $O(P n \log{n})$ operations. The fast algorithm described in this section, in addition to the algorithms described in Section \ref{sec:lit}, further reduces the computational complexity to $O(P n)$. 

  We clarify a few points regarding the FFT algorithm used in this paper. First, while many different FFT algorithms have been developed (e.g. \cite{rabiner1969chirp,good1958interaction}), we focus on the \cite{cooley1965algorithm} algorithm. So whenever we say FFT, we are referring to the Cooley-Tukey algorithm. Next, since the FFT factorizes a length $n$ signal, different algorithms exist for different $n$. This paper only considers in detail when $n$ is a power of two, called the Radix-2 case. However, the Cooley-Tukey algorithm easily extends to arbitrary factorizations. 

  \subsection{Butterflies, Overlapping Trees, and a Fast Algorithm}
  \label{sec:butterfly}
\begin{figure}
    \centering
    \includegraphics[width = 10cm, scale = .5]{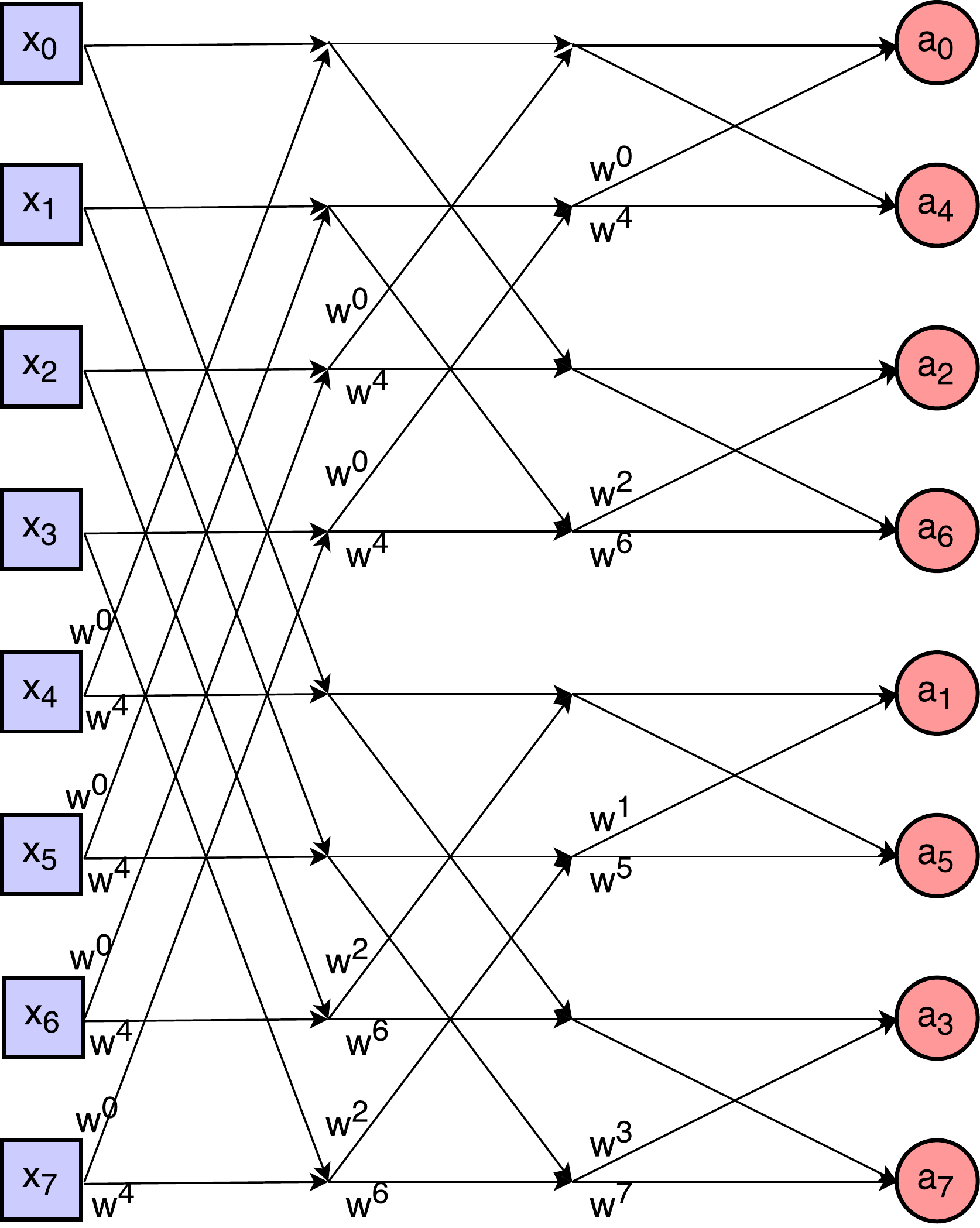}
    \caption{A butterfly diagram for $n = 8$. The squares on the left are the input data, and the circles on the right are the DFT coefficients. Multiplication takes place at the beginning of each arrow, and addition takes place at the end. For example, the first calculation on the bottom left means that $x_7$ is multiplied by $\omega^4$, then added to $x_3$.}
    \label{fig:butterfly} 
  \end{figure}

  The 1D Tree SWDFT algorithm takes a 1D FFT in each window, and avoids repeating calculations already computed in previous windows by storing them in a tree data-structure. Understanding which calculations have already been computed requires a detailed understanding of the FFT. Figure \ref{fig:butterfly} shows the famous butterfly diagram, giving FFT calculations for a length $n = 8$ signal. The squares on the left of Figure \ref{fig:butterfly} correspond to the input data, the circles on the right are the output DFT coefficients, and the arrows in the middle are the calculations. Both \cite{covell1991new} and \cite{farhang1992comment} derived their non-recursive algorithms by showing that calculations in the butterfly diagram repeat in overlapping windows. 

\begin{figure}
    \centering
    \includegraphics[width = 12cm, height = 6cm]{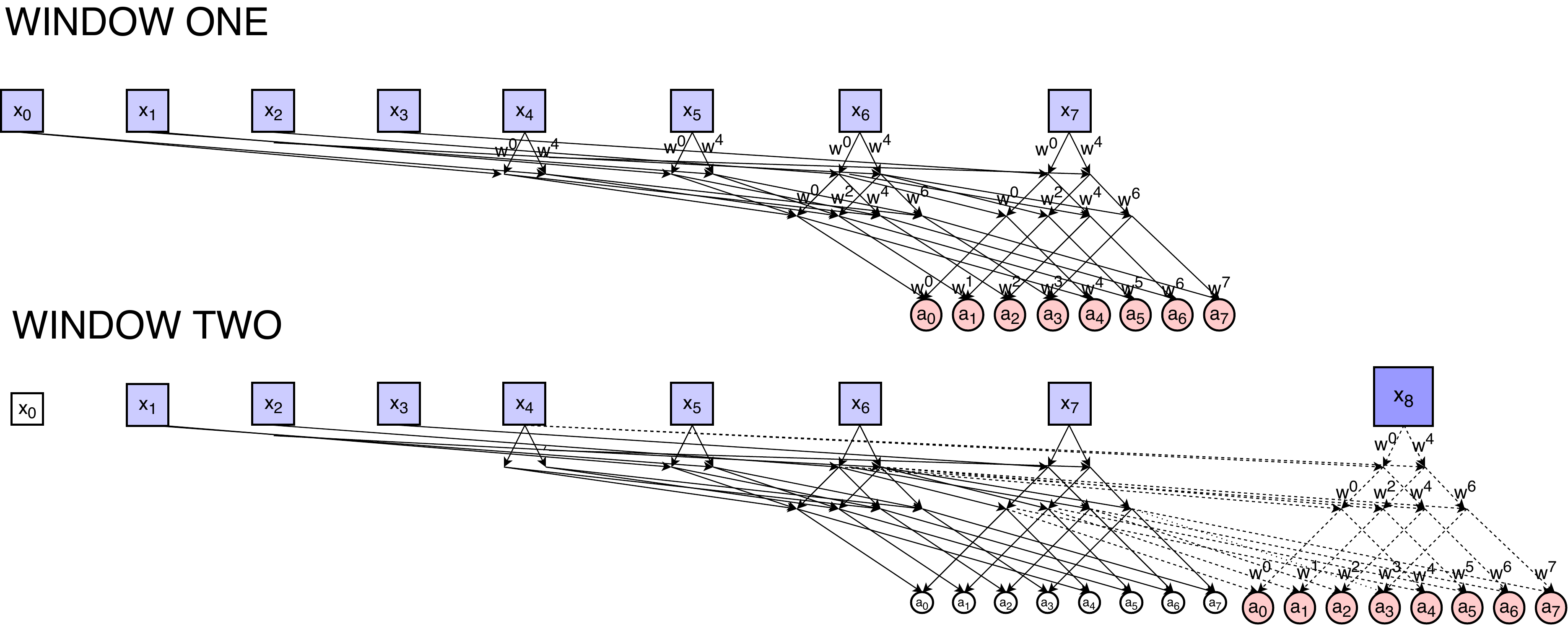}
    \caption{Calculations for the 1D Tree SWDFT algorithm in the first two length $n = 8$ windows. Window one shows the tree data-structure of FFT calculations for input data $[x_0, x_1, \ldots, x_7]$. In window one, the DFT coefficients are circles at the third level of the binary tree underneath $x_7$. Window two shows FFT calculations in window two, where solid arrows indicate window one calculations, and dashed arrows indicate new window two calculations. In window two, the large square indicates the new data-point ($x_8$) entering the window, and the small square indicates the data-point ($x_0$) leaving the window.}
    \label{fig:1dtree}
    \vspace{-1em}
  \end{figure}

  An equivalent tree diagram of the FFT is shown in the top panel of Figure \ref{fig:1dtree}. It is important to stress that the calculations in the butterfly and tree diagrams are the same. For sliding windows, the tree diagram has two major advantages. First, the input data and output coefficients are ordered.  Second, underneath $x_7$ is a binary tree with three (since $2^3 = 8$) levels, and the final level of this binary tree contains the DFT coefficients. 

  Figure \ref{fig:1dtree} demonstrates the 1D Tree SWDFT algorithm. The top panel shows calculations for the window position one with input data $[x_0, x_1, \ldots, x_7]$, and the bottom panel shows the calculations for window position two with input data $[x_1, x_2, \ldots, x_8]$. In window two, solid arrows represent computations made in window one, and dashed arrows represent new window two calculations. The number of new window two calculations is exactly the size of the binary tree, and the difference between the number of solid and dashed arrows is exactly the log factor speed up gained by the 1D Tree SWDFT algorithm. 

  Implementing the 1D Tree SWDFT requires calculating each node of each binary tree, which corresponds to three nested loops:

  \begin{enumerate}
      \setlength\itemsep{.01em}   
      \item over $N$ trees
      \item over $\log_2(n)$ levels
      \item over $2^l$ nodes at level $l$ of each tree.
  \end{enumerate}

  The only restriction on loop order is that loop 2 (over levels) must precede loop 3 (over nodes), since nodes at lower levels of the tree depend on nodes at higher levels. There are three possible loop orders: ($1,2,3$), ($2,1,3$), and ($2,3,1$); the first ordering would be appropriate for data which arrives sequentially in time. 

\section{The 2D Tree Sliding Window Discrete Fourier Transform}
\label{sec:2dfswft}
This section presents the 2D Tree SWDFT algorithm. After defining the 2D SWDFT, we derive the algorithm. We then discuss algorithm implementation, numerical stability, our software package, compare our new algorithm with existing algorithms, and show how the algorithm extends to higher dimensions.

  \subsection{The 2D Sliding Window Discrete Fourier Transform}
  The 2D SWDFT of an $N_0 \times N_1$ array calculates a 2D DFT for all $n_0 \times n_1$ windows. This derivation requires that $n_0 = 2^{m_0}$ and $n_1 = 2^{m_1}$, the Radix-2 case. Let {\bf x} be an $N_0 \times N_1$ array.  There are $P_i = N_i - n_i + 1; i = 0, 1$ window positions in each direction, making $P_0 P_1$ total window positions. Indexing the window position by $(p_0, p_1)$, where $p_i = n_i - 1, n_i, \ldots, N_i - 1; i = 0, 1$, the 2D SWDFT is:

  \vspace{-1em}
  \begin{eqnarray}
    \label{eq:2dswft}
    a_{k_0, k_1}^{p_0, p_1} &=& \frac{1}{\sqrt{n_0 n_1}} \sum_{j_0 = 0}^{n_0 - 1} \sum_{j_1 = 0}^{n_1 - 1} x_{p_0 - n_0 + 1 + j_0, p_1 - n_1 + 1 + j_1} \omega_{n_0}^{-j_0 k_0} \omega_{n_1}^{-j_1 k_1}
  \end{eqnarray}
  \vspace{-1em}

  \hspace{-1.2em} where $k_i = 0, 1, \ldots, n_i - 1; i = 0, 1$. Equation \ref{eq:2dswft} outputs a $P_0 \times P_1 \times n_0 \times n_1$ array.

  A straightforward calculation of Equation \ref{eq:2dswft} takes $P_0 P_1 n_0^2 n_1^2$ operations. Replacing the 2D DFT with a 2D FFT (described next) reduces the number of operations to $O(P_1 P_0 n_0 n_1 \log(n_0 n_1))$. Our algorithm, along with the algorithms of \cite{park20152d} and \cite{byun2016vector}, further reduce this to $O(P_0 P_1 n_0 n_1)$. 

  \subsection{Derivation}
\begin{figure}
    \centering
    \includegraphics[width = 12cm, scale = .5]{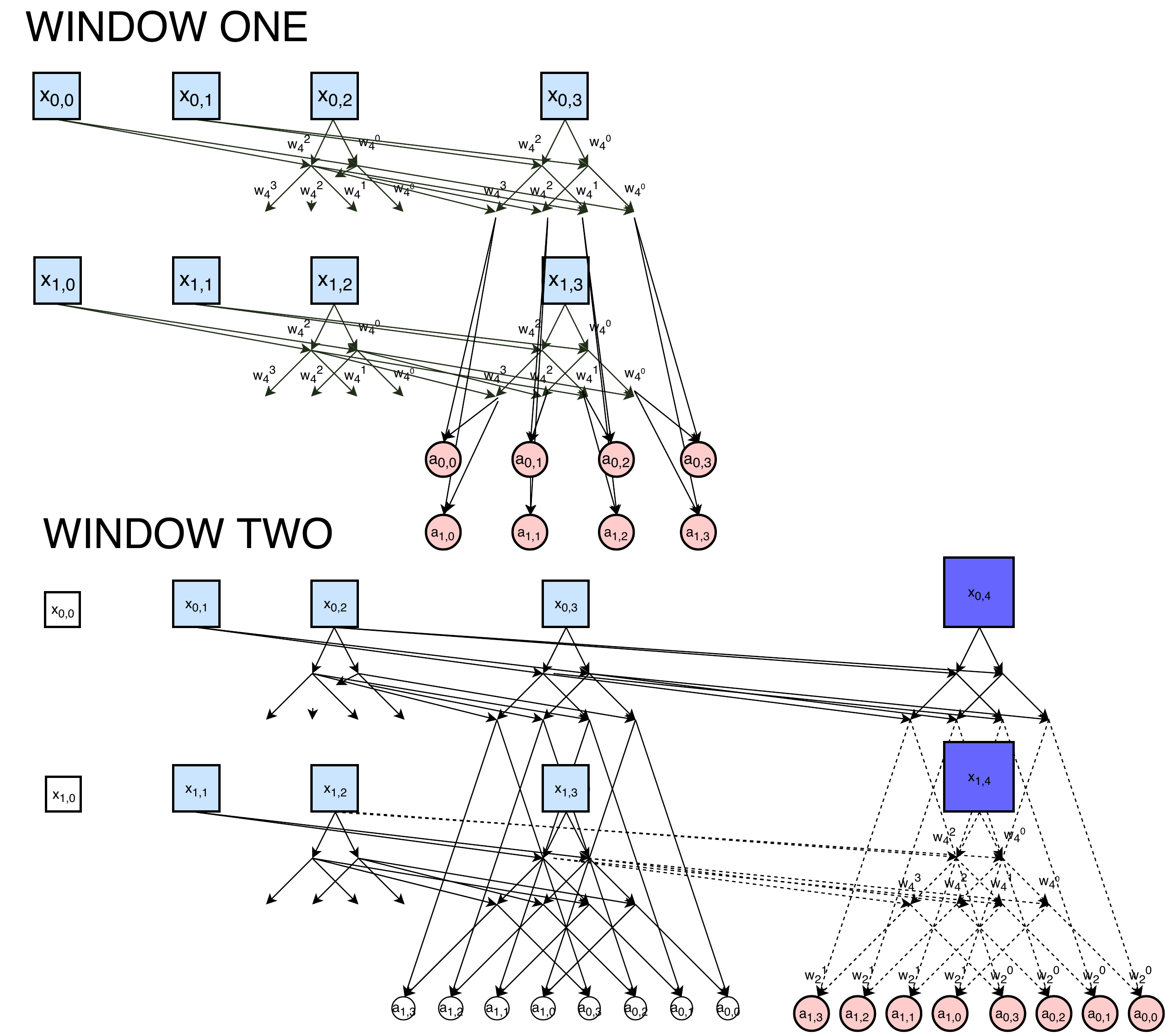}
    \caption{
    The 2D Tree SWDFT for two window positions, where window position one has input data $x_{i,j}, i=0,1, j=0, 1, 2, 3$ and window position two replaces column $0$ ($x_{0, 0}, x_{1, 0}$, the small white squares) with column $4$ ($x_{0, 4}, x_{1, 4}$), the large darker squares). The circles on level three of the 2D binary tree below the bottom-right input data-point of each window position are the 2D DFT coefficients. In the bottom panel, solid lines are calculations from window one, and dashed lines are calculations from window two.}
    \label{fig:2dfft}
  \end{figure}

  Conceptually, the 2D Tree SWDFT algorithm works like 1D. As in 1D, there is a tree data-structure with 2D FFT calculations underneath each data-point (see top panel of Figure \ref{fig:2dfft}). Also like 1D, the 2D Tree SWDFT reuses 2D FFT calculations computed in previous windows. Finally, just like there are different 1D FFT algorithms, there are also different 2D FFT algorithms (see Section 9 of \cite{duhamel1990fast}).

  Since the 2D Tree SWDFT computes a 2D FFT for each window position, and different 2D FFT algorithms exist, our 2D Tree SWDFT derivation requires specifying which 2D FFT to use. The two most commonly used 2D FFTs are the Row-Column FFT and the Vector-Radix FFT. This derivation uses the Row-Column FFT due to its flexibility and straightforward implementation. However, the 2D Tree SWDFT can also be derived by swapping the Row-Column FFT with the Vector-Radix FFT (\cite{byun2016vector}). To be clear, the 2D Tree SWDFT algorithm takes a 2D FFT in each window position, and stores the intermediate calculations in a tree data-structure. The Row-Column FFT is the specific 2D FFT algorithm used in this derivation of the 2D Tree SWDFT algorithm.

  The key to the Row-Column FFT is the ``factorization'' property of the 2D DFT. Factorization means that 2D DFTs can be computed using 1D DFTs: 

  \vspace{-1em}
  \begin{eqnarray}
    a_{k_0, k_1} &=& \sum_{j_0 = 0}^{n_0 - 1} \sum_{j_1 = 0}^{n_1 - 1} x_{j_0, j_1} \omega_{n_0}^{-j_0 k_0} \omega_{n_1}^{-j_1 k_1} \nonumber \\
    &=& \sum_{j_0 = 0}^{n_0 - 1} z_{j_0, k_1} \omega_{n_0}^{-j_0 k_0} 
  \end{eqnarray}
  \vspace{-.5em}

  \noindent For $k_i = 0, \ldots, n_i - 1$, $j_i = 0, \ldots, n_i - 1; i = 0, 1$, where: 

  \begin{eqnarray}
    z_{j_0, k_1} &=& \sum_{j_1 = 0}^{n_1 - 1} x_{j_0, j_1} \omega_{n_1}^{-j_1 k_1}
  \end{eqnarray}

  This implies the 2D DFT can be computed by first taking 1D FFTs of each row ($z_{j_0, k_1}$), followed by 1D FFTs of the resulting columns. This sequence of 1D FFTs is exactly the Row-Column FFT. 

  Though conceptually similar, several details differentiate the 1D and 2D Tree SWDFT algorithms. In 2D, the trees underneath each data-point are two-dimensional, where the two dimensions are $m_0 = \log_2(n_0)$ and $m_1 = \log_2(n_1)$. Second, we now require two twiddle-factor vectors; Twiddle factors (named by \cite{gentleman1966fast}) are the trigonometric constants used for combining smaller DFTs during the FFT algorithm (the $\omega$'s in Figures \ref{fig:butterfly} and \ref{fig:1dtree} are twiddle factors). Finally, when the window slides by one position, we now replace an entire row or column, as opposed to a single data-point in 1D. 

  Figure \ref{fig:2dfft} shows how the 2D Tree SWDFT works. The solid arrows in window two indicate 2D FFT calculations from window one, and the dashed arrows are the new calculations required for window two. Like 1D, the number of dashed lines is exactly the size of the tree underneath the bottom-right point $(x_{1, 4})$ of window two. With this preamble, we are ready to derive the 2D Tree SWDFT. 

  Let {\bf x} be a $N_0 \times N_1$ array, with window sizes $n_0 \times n_1$, where $n_0 = 2^{m_0}$ and $n_1 = 2^{m_1}$, the Radix-2 case. We do not require that $n_0 = n_1$. Our trees have $m_0 + m_1$ levels: the first $m_0$ levels correspond to 1D row FFTs, the next $m_1$ levels correspond to 1D column FFTs, and the final level $(m_0 + m_1)$ is the 2D DFT coefficients. Let $\Omega^{i} = [\omega_{n_i}^{0}, \omega_{n_i}^{-1}, \ldots, \omega_{n_i}^{-(n_i - 1)}]; i = 0, 1$ be a length $n_i$ twiddle-factor vector. Finally, let {\bf T} be the tree data-structure that stores 2D FFT calculations for each window position. We access {\bf T} by window position, level, and node. For example, $T_{p_0, p_1, l, i_0, i_1}$ corresponds to node $(i_0, i_1)$ on the $l^{th}$ level of the tree at window position $(p_0, p_1)$. 

  We give the calculations for an arbitrary tree at window position $(p_0, p_1)$. Each level $l$ has $2^{l}$ nodes. Level zero of the tree is the data: $T_{p_0, p_1, 0, 0, 0} = x_{p_0, p_1}$. We use one additional piece of notation: let $s_{l}^{d}$ be the ``shift'' for dimension $d$ and level $l$. The ``shift'' ($s_l^d$) identifies which tree has the repeated calculation needed for the current calculation. For example, if $s_{1}^{0} = 2$, this means for level 1, the repeated calculation is located at tree $(p_0 - 2, p_1)$. For levels corresponding to row FFTs: $s_{l}^{1} = 2^{m_1 - l}$, and for column FFT levels: $s_{l}^{0} = 2^{m_0 + m_1 - l}$.

  Level one of tree $(p_0, p_1)$ has $2^1 = 2$ nodes: $(0, 0), (0, 1)$. The shift is $s_{1}^{1} = 2^{m_1 - 1}$, meaning the repeated calculation is at tree $(p_0, p_1 - s_{1}^{1})$. For node $(0, i_1)$, the calculation is a complex multiplication between node $(0, i_1 \text{ mod } 2^{1 - 1}) = (0, 0)$ at level $1 - 1 = 0$ of the current tree $(p_0, p_1)$, and the $i_1 \cdot s_{1}^{1}$ element from twiddle-factor vector $\Omega^1$. The complex multiplication output is then added to node $(0, 0)$ from the shifted tree $(p_0, p_1 - s_{1}^{1})$. Using our tree notation:

  \vspace{-1em}
  \begin{eqnarray}
  T_{p_0, p_1, 1, 0, i_1} &=& \nonumber T_{p_0, p_1 - s_{1}^{1}, 0, 0, 0} + \Omega^1_{i_1 \cdot s_{1}^{1}} \cdot T_{p_0, p_1, 0, 0, 0} \\
  \nonumber
  \end{eqnarray}
  \vspace{-2em}

  \hspace{-1.2em} for $i_1 = 0, 1$, which is a length two DFT of $x_{p_0, p_1}$ and $x_{p_0, p_1 - s_{1}^{1}}$.

  Calculating each node of {\bf T} has the same form: $T_A = T_B + \Omega^i_C \cdot T_D$. $A, B, C$ and $D$ are indices into either {\bf T}, $\Omega^0$, or $\Omega^1$, depending on whether the level $l$ of the tree corresponds to the row ($1$ to $m_1$) or column ($m_1 + 1$ to $m_1 + m_0$) part of the Row-Column FFT. We give the exact calculations for both situations next. 

  Level $t \leq m_1$ of tree $(p_0, p_1)$ has $2^t$ nodes: 1 in the row-direction, and $2^t$ in the column-direction. The repeated calculation comes from tree $(p_0, p_1 - s_{t}^1)$, where $s_{t}^1 = 2^{m_1 - t}$. For node $(0, i_1); i_1 = 0, 1, \ldots 2^{t} - 1$, the calculation is:

  \vspace{-1em}
  \begin{eqnarray}
  \label{eq:rowfswft2d}
  T_{p_0, p_1, t, 0, i_1} &=& T_{p_0, p_1 - s_{t}^{1}, t - 1, 0, i_1 \text{ mod } 2^{t - 1}} + \Omega^1_{i_1 \cdot s_t^1} \cdot T_{p_0, p_1, t - 1, 0, i_1 \text{ mod } 2^{t - 1}} \\
  \nonumber
  \end{eqnarray}
  \vspace{-2em}

  Level $v > m_1$ of tree $(p_0, p_1)$ has $2^v$ nodes: $2^{v - m_1}$ in the row-direction, and $n_1$ in the column-direction. The repeated calculation comes from tree $(p_0 - s_v^0, p_1)$, where $s_{v}^{0} = 2^{m_0 + m_1 - l}$. For node $(i_0, i_1); i_0 = 0, 1, \ldots n_1, i_1 = 0, 1, \ldots 2^{v - m_1} - 1$, the calculation is:

  \vspace{-1em}
  \begin{eqnarray}  
  \label{eq:colfswft2d}
  T_{p_0, p_1, v, i_0, i_1} &=& T_{p_0 - s_v^0, p_1, v - 1, i_0 \text{ mod } 2^{v - m_1 - 1}, i_1} + \Omega^0_{i_0 \cdot s_{v}^0} \cdot T_{p_0, p_1, v - 1, i_0 \text{ mod } 2^{v - m_1 - 1}, i_1} \\
  \nonumber
  \end{eqnarray}  
  \vspace{-2em}

  The final level $(m_0 + m_1)$ contains the 2D DFT coefficients for window position $(p_0, p_1)$. After calculating {\bf T}, all that remains is selecting the subset at level $m_0 + m_1$ of each tree, and the algorithm is complete. 

  \subsection{Algorithm Implementation}
  After creating the twiddle-factor vectors and allocating memory, we implement the 2D Tree SWDFT in six nested loops: 

  \vspace{-.2em}
  \begin{enumerate}
    \setlength\itemsep{.01em}   
    \item over $m_0 = \log_2(n_0)$ levels corresponding to row FFTs 
    \item over $m_1 = \log_2(n_1)$ levels corresponding to column FFTs 
    \item over $N_0$ trees in the row-direction 
    \item over $N_1$ trees in the column-direction 
    \item over $2^{l_0}$ nodes in the row-direction at level $l_0$
    \item over $2^{l_1}$ nodes in the column-direction at level $l_1$
  \end{enumerate}

  Inside the six loops is either Equation \ref{eq:rowfswft2d} or \ref{eq:colfswft2d}, depending on whether the level corresponds to the row or column part of the Row-Column FFT algorithm.

  Since the next level of the trees only depends on the previous level, our implementation allocates  $2 N_0 N_1 n_0 n_1$ complex numbers in memory: $N_0 N_0 n_0 n_1$ for both the previous and current level. This is possible because the two loops over levels are innermost. This also implies our algorithm is suitable for parallel computing, following \cite{wang2012parallel}. 

  Like 1D, we can swap the order of the loops. For our derivation, the only restrictions are that loop $1$ must precede loop $2$, and loops $5$ and $6$ must come after loops $1$ and $2$. A particularly interesting order is when loops $3$ and $4$ (over trees) are innermost. This version can be tailored to a real-time task, opposed to the ``levels innermost'' order, which requires all data to be available. For example, if the channels of a multi-spectral scanner are considered as a two-dimensional array, and {\bf T} is stored, we could calculate the output for a new window in $O(n_0 n_1)$ time. 

  The 2D SWDFT is memory intensive, since the output is a $P_0 \times P_1 \times n_0 \times n_1$ array. For example, if $P_0 = P_1 = 400$, and $n_0 = n_1 = 32$, then the output is $163,840,000$ complex numbers. Because our implementation uses $16$ byte double complex numbers, the output alone requires $16 \cdot 163840000 = 2621440000$ bytes, or $\approx 2.6$ gigabytes. 

\subsection{Extension to $k$D}
  Our algorithm extends directly to $k$D. For instance, if we had an $N_0 \times N_1 \times \ldots \times N_{k - 1}$ array, with $n_i = 2^{m_i} \leq N_i;i = 0, \ldots, k - 1$  windows, then the calculation for node $(i_0, i_1, \ldots, i_{k - 1})$, level $l$, and dimension $c$ is:

  \vspace{-1em}
  \begin{eqnarray}
    \hspace{-2em}
    \label{eq:kdfswft}
    T_{p_0,.., p_{k - 1}, l, 0,\ldots,i_{c}, i_{c + 1},\ldots,i_{k - 1}} &=& T_{p_0,.., p_c - s_l^c,.., p_{k - 1}, l - 1, 0,.., i_c \text{ mod } 2^{l - 1 - \sum_{i = c + 1}^{k - 1} m_i},i_{c + 1},..,i_{k - 1}} \nonumber \\ 
    &+& \Omega^c_{i_c \cdot s_l^c} \cdot T_{p_0, \ldots, p_{k - 1}, l - 1, 0, .., i_c \text{ mod } 2^{l - 1 - \sum_{i = c + 1}^{k - 1} m_i}, i_{c + 1},.., i_{k - 1}} \\
    \nonumber
  \end{eqnarray}  

  \hspace{-1.2em} where $s_{l}^{c} = 2^{(\sum_{i = c}^{k - 1} m_i) - l}$, and dimension $c$ contains levels $[\sum_{i = c + 1}^{k - 1} m_i + 1, \sum_{i = c}^{k - 1} m_i]$ of the trees. Extending the implementation to any dimension is possible following Equation \ref{eq:kdfswft}. For example, the $3$D implementation takes $9$ loops: $3$ over dimension, $3$ over trees, and $3$ over nodes. The computation inside the loops is one of:

  \begin{eqnarray}
    T_{p_0, p_1, p_2, l, 0, 0, i_2} &=& T_{p_0, p_1, p_2 - s_l^2, l - 1, 0, 0, i_2 \text{ mod } 2^{l - 1}} + \Omega^2_{i_2 \cdot s_l^2} \cdot T_{p_0, p_1, p_2, l - 1, 0, 0, i_2 \text{ mod } 2^{l - 1}} \nonumber \\
    T_{p_0, p_1, p_2, l, 0, i_1, i_2} &=& T_{p_0, p_1 - s^1_l, p_2, l - 1, 0, i_1 \text{ mod } 2^{l - m_2 - 1}, i_2}  + \Omega^1_{i_1 \cdot s_l^1} \cdot T_{p_0, p_1, p_2, l - 1, 0, i_1 \text{ mod } 2^{l - m_2 - 1}, i_2} \nonumber \\
    T_{p_0, p_1, p_2, l, i_0, i_1, i_2} &=& T_{p_0 - s_l^0, p_1, p_2, l - 1, i_0 \text{ mod } 2^{l - m_1 - m_2 - 1}, i_1, i_2} + \Omega^0_{i_0 \cdot s_l^0} \cdot T_{p_0, p_1, p_2, l - 1, i_0 \text{ mod } 2^{l - m_1 - m_2 - 1}, i_1, i_2} \nonumber 
  \end{eqnarray}

  Depending on if level $l$ of the tree is between $[1, m_2]$, $[m_2 + 1, m_2 + m_1]$, or $[m_2 + m_1 + 1, m_2 + m_1 + m_0]$, respectively. The $4$D implementation follows the same idea with $3 \cdot 4 = 12$ loops, and so on. 

  \subsection{Software Description}
  In the software package accompanying this paper, our primary contribution is a C implementation of the 2D Tree SWDFT algorithm. The purpose of our implementation is not a specially optimized program, but rather a precise description of how the algorithm works. 

\begin{figure}[ht]
    \centering
    \includegraphics[width = 10cm]{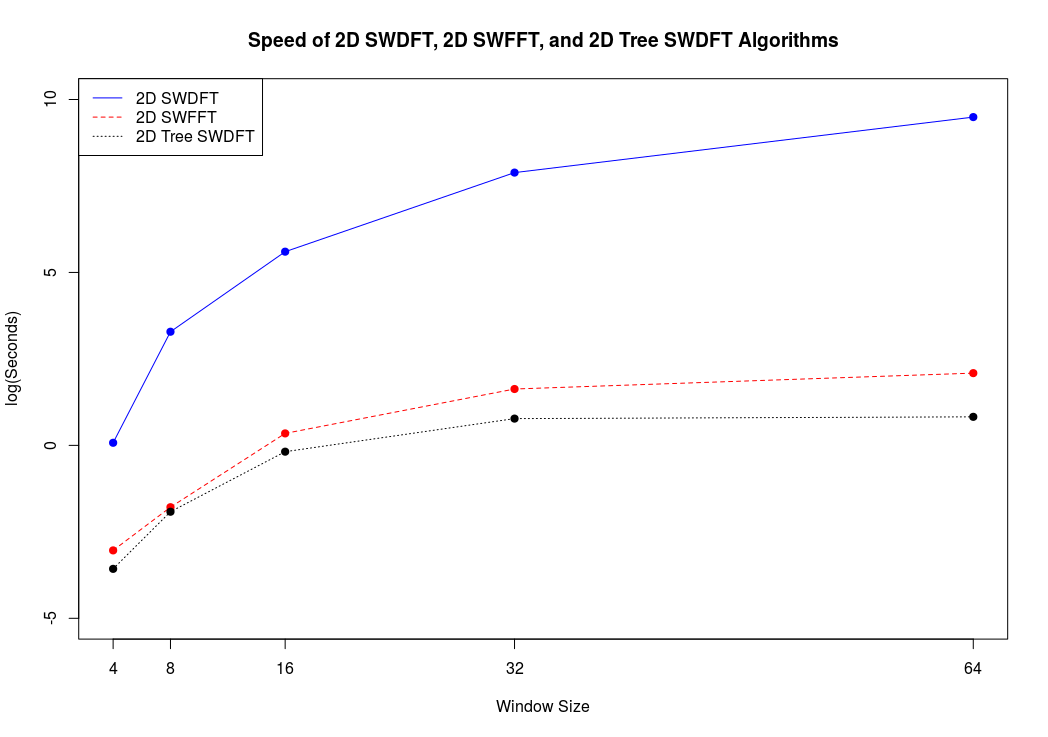} 
    \caption{Speed comparison of the 2D SWDFT, 2D SWFFT, and 2D Tree SWDFT algorithms. Each algorithm is run on an $100 \times 100$ array, with window sizes $n_0 = n_1 = 4, 8, 16, 32$, and $64$. This shows the speedup due to fewer operations in the 2D Tree SWDFT, controlling for the implementation details.}
    \label{fig:speed-compare}
  \end{figure}

  Our package is small and light-weight, and only requires a {\bf gcc} compiler and C standard libraries to run. The only function included in our primary source code is the 2D Tree SWDFT implementation. We include two other functions to verify the accuracy and stability of our implementation: the 2D SWDFT and 2D SWFFT. The 2D SWDFT takes a 2D DFT in each window position, and the 2D SWFFT takes a 2D FFT in each window position. We include these two functions because other algorithms of comparable theoretical speed (\cite{park20152d,byun2016vector}), to our knowledge, do not have publicly available software implementations. Figure \ref{fig:speed-compare} compares the speed of our 2D SWDFT, 2D SWFFT, and 2D Tree SWDFT implementations. 

  \subsection{Numerical Stability}
  The initial motivation for non-recursive SWDFT algorithms was numerical stability. As discussed in Section \ref{sec:lit}, recursive algorithms were proposed before non-recursive algorithms, but were quickly identified as numerically unstable. The reason recursive algorithms are numerically unstable is that the complex exponentials (e.g. $\omega_n$) are represented with finite precision, and the rounding errors accumulate as the number of window positions increases. In contrast, non-recursive algorithms use the same underlying calculations as FFT algorithms, so the output is identical to the 2D SWFFT. Another way to say this is that the the number of operations for non-recursive algorithms is bounded. For example, computing each coefficient with the 2D Tree SWDFT algorithm takes exactly $2(n_0 n_1 - 1)$ operations (see Equation \ref{eq:windowops} below). We include a program in the tests of our software package that shows our algorithm gives identical results to the 2D SWFFT, since both algorithms use identical intermediate calculations. 

  \subsection{Results and Comparisons}
  The run-time of the 2D Tree SWDFT algorithm grows linearly in window ($n_0 n_1$) and array ($N_0 N_1$) size. This is because calculating each node requires one complex multiplication and one complex addition, defined as an operation (the same definition was used in \cite{cooley1965algorithm}). Since level $l$ requires $2^l$ operations, number of operations per window is:

  \vspace{-1em}
  \begin{eqnarray}
    \label{eq:windowops}
    C_{each} &=& \sum_{l = 1}^{m_0 + m_1} 2^l = (2 + 4 + \ldots + 2^{m_0 + m_1}) = 2 (2^{m_0 + m_1} - 1) = 2 (n_0 n_1 - 1) 
  \end{eqnarray}  

  The exact run-time is slightly more complicated, since windows with indices either less than $n_0$ in the row direction or less than $n_1$ in the column direction do not require complete trees. These extra calculations are negligible for large arrays, so the 2D Tree SWDFT takes $O(P_0 P_1 n_0 n_1)$ operations. Table \ref{tab:fswft2d_compare} compares the speed, memory, and properties of the 2D Tree SWDFT with existing algorithms. Our memory numbers for \cite{park20152d} and \cite{byun2016vector} come from the Table 2 in \cite{byun2016vector}. Out of existing $O(P_0 P_1 n_0 n_1)$ algorithms, our 2D Tree SWDFT is the only one that is numerically stable, works for non-square windows, and has an existing publicly available software implementation. 

  \begin{table}[ht]
    \centering 
    \scalebox{.8}{
    \begin{tabular}{|c|c|c|c|p{24mm}|}
      \hline
      {\bf Algorithm} & {\bf Speed} & {\bf Memory} & {\bf Numerically Stable} & {\bf Non-Square Windows} \\
      \hline 
      \hline 
      2D SWDFT                                    & $O(P_0 P_1 n_0^2 n_1^2)$           & $P_0 P_1 n_0 n_1$                           & Yes & Yes \\ \hline
      2D SWFFT                                    & $O(P_0 P_1 n_1 n_0 \log(n_0 n_1))$ & $P_0 P_1 n_0 n_1$                           & Yes & Yes \\ \hline
      \cite{park20152d}                           & $O(P_0 P_1 n_0 n_1)$               & $P_0 P_1 n_0 n_1 + 2 n + 1$                 & No  & Yes \\ \hline
      \cite{byun2016vector}                 & $O(P_0 P_1 n_0 n_1)$               & $P_0 P_1 n_0 n_1 + n^2 2 \log_2(n - 1) + 1$ & Yes & No  \\ \hline      
      This Paper                              & $O(P_0 P_1 n_0 n_1)$               & $2 N_0 N_1 n_0 n_1$                         & Yes & Yes \\ 
      \hline   
    \end{tabular}}    
    \caption{Speed, memory, and properties of existing 2D SWDFT algorithms. The 2D SWDFT and SWFFTs take a 2D DFT or 2D FFT in each window. \cite{park20152d} extends the 1D recursive algorithm to 2D, and \cite{byun2016vector} gives a Vector-Radix $2 \times 2$ algorithm. The memory results come from \cite{byun2016vector}, and we use $n$ instead of $n_0 n_1$, because the algorithms in this paper were derived using non-square window sizes. }
    \label{tab:fswft2d_compare} 
  \end{table}

  \section{Discussion}
  \label{sec:discussion}
  The goal of this paper is to describe our 2D SWDFT algorithm as clearly as we can, and instantiate it with a C program. 
  
  Finally, while our paper focused on the Radix-2 case, nothing conceptually prevents extension to other factorizations. For instance, the Radix-3 implementation simply replaces binary trees with ternary trees. For higher dimensions, we simply replace Equations \ref{eq:rowfswft2d} and \ref{eq:colfswft2d} with Equation \ref{eq:kdfswft}. 

  \appendix 
  \section{Software Availability}
  The implementation of our algorithm is available online at: \url{http://stat.cmu.edu/~lrichard/xxx.tar.gz}

\clearpage 
\bibliography{references}

\end{document}